\documentclass[aps,preprint,floatfix,bibnotes,altaffilletter]{revtex4}

\usepackage{dcolumn}
\usepackage[usenames,dvipsnames]{color}
\usepackage{amsmath}
\usepackage{here}
\usepackage[dvips]{graphicx}

\newcommand\pictc[5]{\begin{figure}
                       \centerline{
                       \includegraphics[width=#1\columnwidth]{#3}}
                   \protect\caption{\protect\label{fig:#4} #5}
                    \end{figure}            }
\newcommand\pict[4][1.]{\pictc{#1}{!tb}{#2}{#3}{#4}}
\newcommand\rpict[1]{\ref{fig:#1}}

\newcommand{\be}{\begin{equation}}
\newcommand{\ee}{\end{equation}}
\newcommand{\bea}{\begin{eqnarray}}
\newcommand{\eea}{\end{eqnarray}}

\renewcommand{\abstractname}{}\abstractname

\newcounter{Fig}

\begin{document}

\begin{sloppy}

\title{Electromagnetic wave analogue of electronic diode}

\author{Ilya V. Shadrivov$^*$}

\affiliation{Nonlinear Physics Center, Research School of Physics and
Engineering, Australian National University, Canberra ACT 0200, Australia}

\author{David A. Powell}

\affiliation{Nonlinear Physics Center, Research School of Physics and
Engineering, Australian National University, Canberra ACT 0200, Australia}

\author{Vassili A. Fedotov}

\affiliation{Optoelectronics Research Centre and Centre for Photonic Metametarials, University of Southampton, SO17 1BJ, United Kingdom}

\author{Nikolay I. Zheludev}

\affiliation{Optoelectronics Research Centre and Centre for Photonic Metametarials, University of Southampton, SO17 1BJ, United Kingdom}

\author{Yuri S. Kivshar}

\affiliation{Nonlinear Physics Center, Research School of Physics and
Engineering, Australian National University, Canberra ACT 0200, Australia}

\begin{abstract}
\vspace{20pt}
{\bf An electronic diode is a nonlinear semiconductor circuit component that allows conduction of electrical current in one direction only. A component with similar functionality for electromagnetic waves, an electromagnetic isolator, is based on the Faraday effect of the polarization state rotation and is also a key component of optical and microwave systems. Here we demonstrate a chiral electromagnetic diode, which is a direct analogue of an electronic diode: its functionality is underpinned by an extraordinary strong nonlinear wave propagation effect in the same way as electronic diode function is provided by a nonlinear current characteristic of a semiconductor junction. The effect exploited in this new electromagnetic diode is an intensity-dependent polarization change in an artificial chiral metamolecule. This microwave effect exceeds a similar optical effect previously observed in natural crystals by more than 12 orders of magnitude and a direction-dependent transmission that differing by a factor of 65.}
\end{abstract}

%\pacs{81.05.Xj, 42.65.Pc, 78.20.Ek, 41.20.Jb}
\maketitle

Lorentz reciprocity Lemma dictates that \emph{linear} transmission of certain polarization states must be identical for forward and backward directions, unless the media is statically magnetized or causes polarization conversion. In optics, the simplest isolator exploiting nonreciprocal transmission of circularly polarized light consists of a pair of polarizers and a Faraday rotator and it requires a static magnetic field~\cite{Optics_Encyclopedia}. A similar approach is also used for microwave devices. As another example, asymmetric transmission in metamaterial structures is allowed if propagation is accompanied by polarization conversion~\cite{ASS,Menzel:PRL}.

The Lorentz Lemma is not applicable to nonlinear effects, thus allowing asymmetric transmission of intense electromagnetic waves. Asymmetric nonlinear distributed Bragg gratings~\cite{Tocci:1995-2324:APL}, metamaterial structures~\cite{Feise:2005-37602:PRE}, and even disordered layered structures~\cite{Shadrivov:2010-123902:PRL} were predicted theoretically to show directionally asymmetric response for linearly polarized light. In this Letter, we introduce and verify experimentally the concept of {\em a nonlinear electromagnetic diode for circularly polarized waves}. It is analogous to an electronic diode that transmits electric current in only one direction due to its nonlinear current-voltage characteristics, see Fig.~\rpict{principle}. The nonlinear element in our nonreciprocal structure is an {\em artificial chiral metamolecule}.  By introducing nonlinearity into the metamolecule we experimentally demonstrate that it exhibit unidirectional transmission for one circular polarization whilst remaining transparent for the polarization of opposite handedness (see Fig.~\rpict{principle}).

%%%%%%%%%%%%%%%%
\pict[0.6]{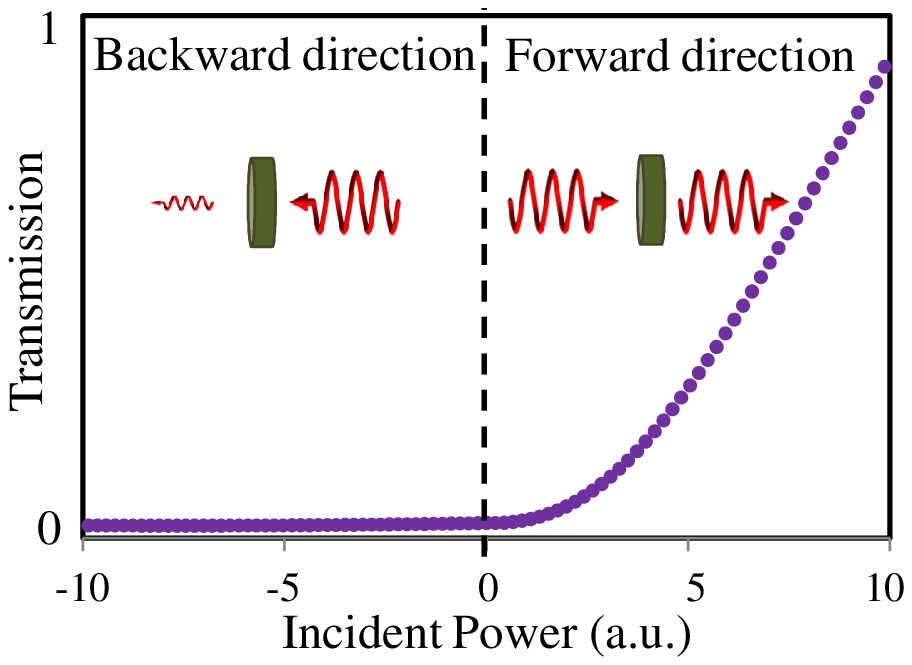}{principle}{{\bf Operating principles of a metamaterial diode.} The device shows different levels of transmission for circularly polarized waves propagating in the opposite directions due to the nonlinearity of a chiral metamolecule. This is similar to an electronic diode exhibiting different resistance for currents propagating in the opposite directions
due to nonlinearity of the p-n junction.}
%%%%%%%%%%%%%%%%

Nonlinear asymmetric transmission can only occur in the presence of a strong intensity-dependent propagation effect. Our idea is to exploit the intensity dependence of the gyrotropy in chiral media, which manifests itself as differential circular birefringence and dichroism. It has been known for nearly two hundred years that many natural media exhibit strong optical activity, however theoretical estimates predicted that nonlinear optical activity would be small and difficult to observe ~\cite{Akhmanov,Kielich,Atkins}. The first mention of this effect was made by S.I. Vavilov~\cite{Vavilov}, who concluded that the necessary light intensities could only be found inside the stars. The subsequent invention of the laser made it possible to study this effect with experimentally achievable power levels, and it was first observed in 1979 in $LiIO_3$ crystals~\cite{Akhmanov:1979-294:JETPL}. In such materials the nonlinear optical activity was smaller than its linear counterpart by a factor of $10^{-6}$, and this required samples several centimeters in length and light intensities of $100$~MW/cm$^2$, which is close to the optical breakdown of the crystal. Such a small level of nonlinearity is not sufficient for demonstrating any practically important electromagnetic diode functionality. In artificial meta-molecules, however, much higher levels of optical activity can be achieved through the engineering of chiral properties. For instance, the polarization rotation in microwave chiral metamaterials can be nearly a million times stronger than in natural quartz for optical frequencies, once the sample thickness is normalized to the wavelength of radiation~\cite{Rogacheva:2006-177401:PRL}. Moreover, achieving high levels of nonlinearity in metamaterials is much easier than in natural media through the use of nonlinear electronic components integrated into metamaterial structures for microwave applications\cite{Shadrivov:2008-161903:APL,Powell:2009-084102:APL}, or by exploiting the local-field enhancement effects in the optical range~\cite{CNT}. This leads to the opportunity to observe {\em extremely strong nonlinear gyrotropy} and thus to develop an electromagnetic diode.

We create the microwave metamaterial diode with a nonlinear chiral metamolecule constituted by a pair of wire strips, which are separated by a dielectric layer and twisted with respect to each other by angle $\phi$, as shown in the inset to Fig. 2(a) (see Methods for more details). The resonant mode of this structure interacts with both the electric and magnetic fields due to the inductive chiral coupling between the wires, resulting in strong gyrotropy~\cite{Svirko:2001-498:APL}. Adding a nonlinear element, a lumped varactor, to the center of one of the wires allows for the creation of the metamaterial diode with high-contrast asymmetry in forward and backward transmission in the high-power regime -- see schematics in Fig.~\rpict{diode}, and the more detailed description in Methods. %We also investigate metamolecules containing nonlinear elements at the center of both wires and observe a profound bistable response.
%We expect that our ideas can be extended to the applications in optical frequency range.

%%%%%%%%%%%%%%%%
\pict[0.6]{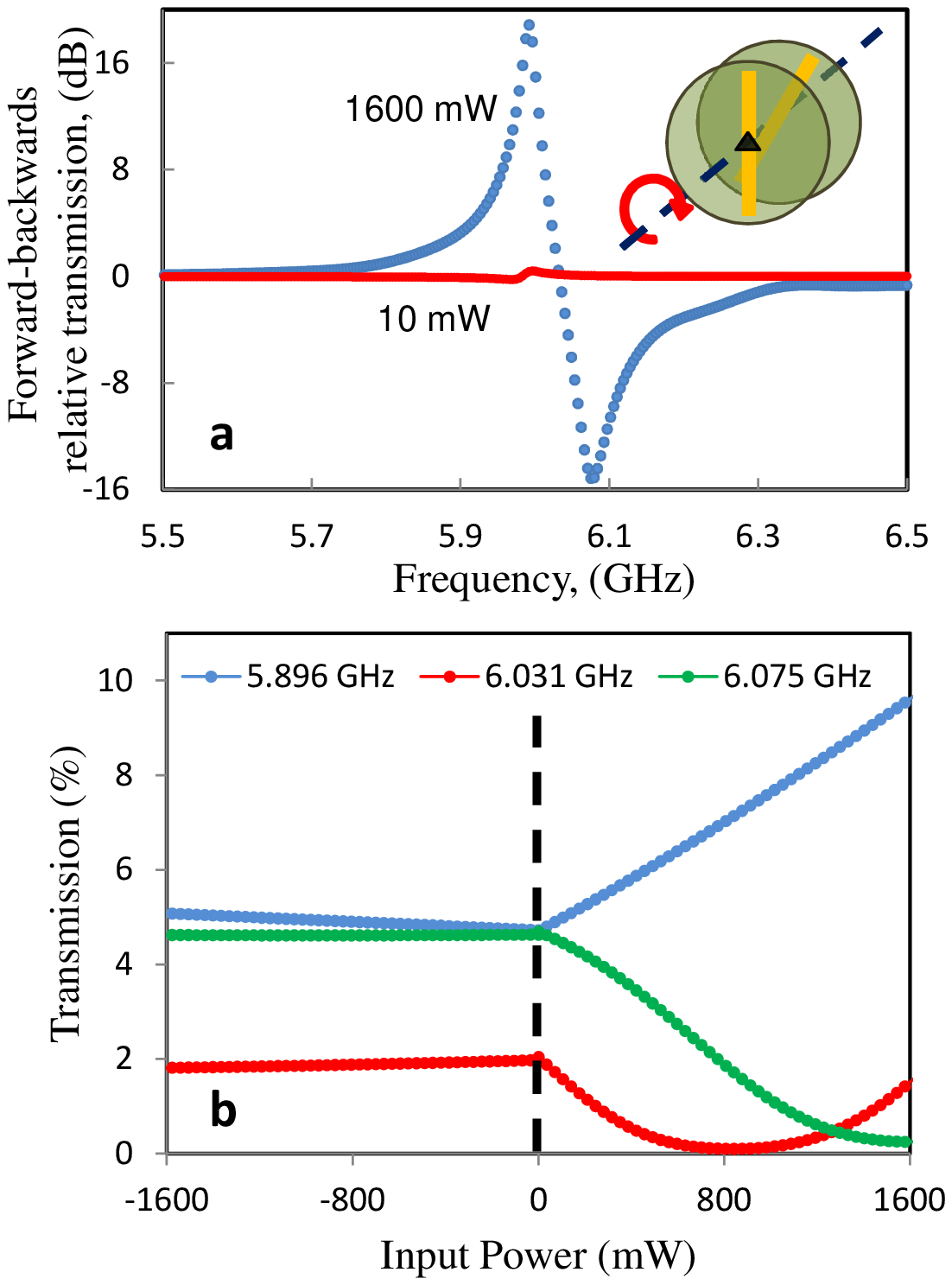}{diode}{{\bf Asymmetric transmission properties of a single chiral metamolecule. } {\bf a:} ratio of the transmission coefficients in forward and backward directions (in dB). For low power (red curve), the transmission is equivalent in both directions for left-handed circularly polarized waves. For large power (blue curve), the transmission is strongly asymmetric. It is larger in the forward direction for the frequencies below 5.97 GHz, and it is larger in backward direction for higher frequencies. The difference reaches 18 dB, or a factor of approximately 65. Inset shows schematic of the metamolecule. {\bf b:} Transmission as a function of incident power. Positive incident power corresponds to the forward propagation, while negative values describe backward propagation. Depending on the frequency, the diode can change polarity, so that it is 'open' in the forward direction (blue curve) or in the backward direction (green curve). Additionally, the diode can suppress transmission in a certain range being open only for higher or lower wave intensities (red curve). The spectra were obtained using time gating.}
%%%%%%%%%%%%%%%%

Results of our measurements for the left-handed circularly polarized wave scattering on left-handed chiral metamolecule are shown in Fig.~\rpict{diode}. When the amplitude of the incident wave is small, the structure shows a linear response, and the transmission coefficients in both directions are equal. However, in the nonlinear regime, with high intensity of the impinging wave we observe considerably different transmission properties in opposite directions with the maximal intensity contrast between the two directions of 18 dB. Our numerical modeling shows that such behavior results from significantly different current amplitudes induced in the two wire strips by the waves entering the metamolecule from one direction, in comparisons with a much smaller excitation difference produced by a wave entering from the opposite direction. The 'polarity' of the metamaterial diode depends on the operating frequency: In the range 5.9 - 6.0 GHz the transmission for the left-handed circularly polarized wave is greater in the forward direction (see Fig. 2(a)), i.e. when the wave hits the strip with the nonlinear element first.  However, in the range from 6.0 GHz to 6.3 GHz the 'polarity' reverses and the diode transmits the same polarization in the opposite direction only.

Figure~\rpict{diode} (b) demonstrates the dependence of the transmittance on the incident power in different directions. The transmission curves are asymmetric with respect to zero incident power indicating a remarkable similarity with the I-V response of an electronic diode, while switching the 'polarity' depending on the frequency of operation. We note that the response time of the structure is below 1 sec for increasing incident power, and it is of the order of 10 sec for decreasing microwave power. The nature of nonlinear response of the diode in resonant system was discussed before in Ref.~\cite{Powell:2007-144107:APL}, and it is caused by the rectification of the alternating current on the diode, which induces self-bias voltage.
%Note that moderate overall transmission coefficient at the resonance is caused by strong absorption in the dielectric substrate of the metamolecule and its nonlinear element.

To study the nonlinear polarization properties of the chiral metamolecule, we also measured its polarization rotatory power for the symmetric case, when central sections of both wires are loaded with varactors. We notice that the angle of twist is an important parameter, which not only determines the magnitude of the gyrotropy~\cite{Svirko:2001-498:APL,Rogacheva:2006-177401:PRL}, but also changes the resonant frequency of the metamolecule due to strong near-field interactions (see Ref.~\cite{Lapine:2009-084105:APL} and references therein).

In general chiral and anisotropic structures have elliptically polarized eigenmodes, leading to a complex dependence of the polarization state of transmitted wave on the incident polarization state. We observed no dependence of our results on the orientation of the metamolecule when it was rotated along its axis in the cylindrical waveguide. This indicates that anisotropic, birefringence effects are negligible and the polarization change is dominated by circular dichroism and circular birefringence of the sample. This allows us to describe the transmission in terms of the transmission of the left- and right-handed circularly polarized waves, $T_{--}$ and $T_{++}$.

%%%%%%%%%%%%%%%%%
\pict[0.7]{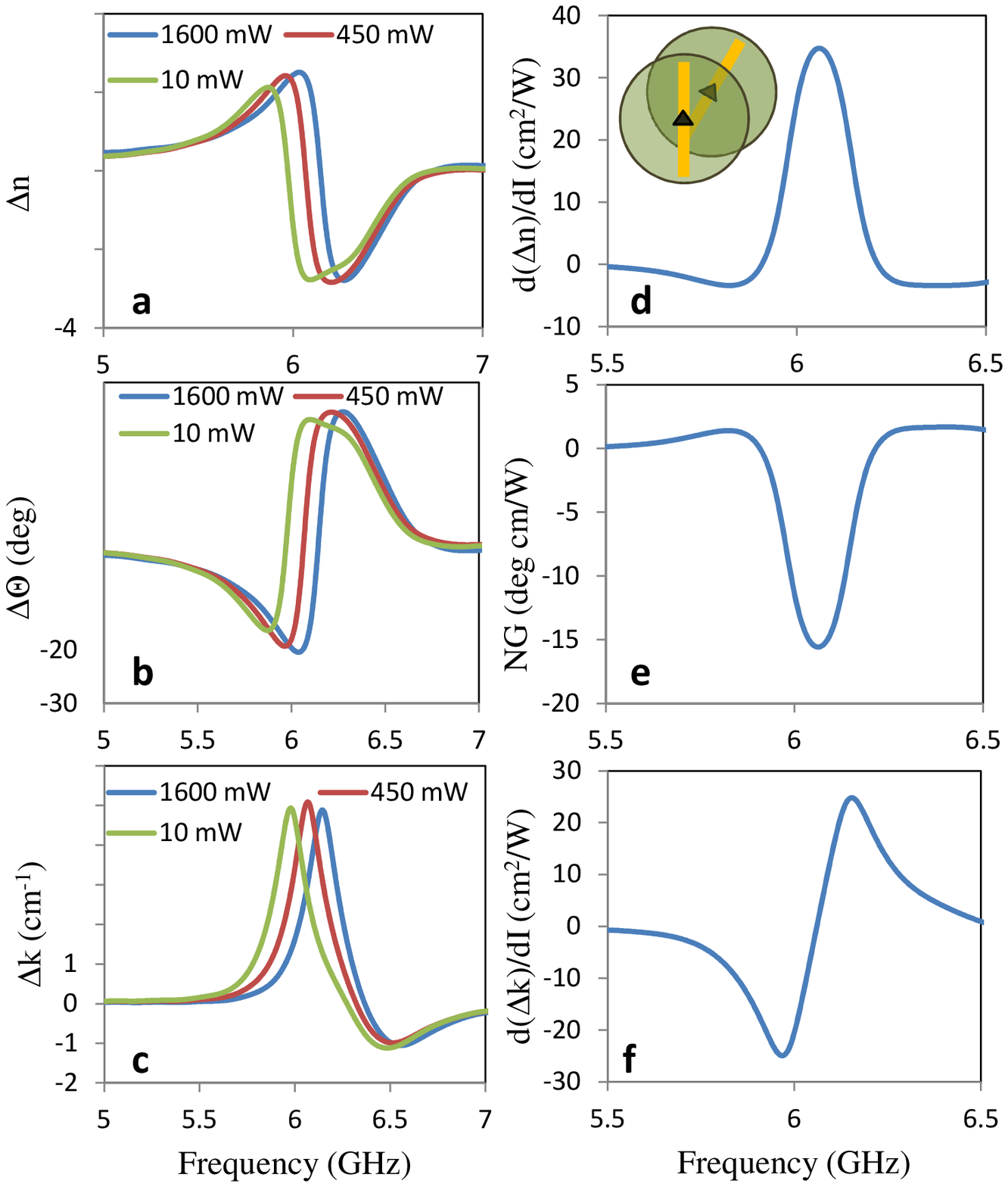}{properties}{{\bf Giant nonlinear polarization rotation in a symmetric chiral meta-molecule.} (a,b,c) Circular birefringence $n_+ - n_-$, angle of polarization rotation, and wavenumber difference vs. frequency for several values of incident power. (d,e,f) nonlinear gyrotropy shown as power-driven variation of (d) circular birefringence, (e) polarization azimuth rotation, and (f) wavenumber difference. The inset shows a schematic of the left-handed chiral metamolecule composed of twisted wires identically loaded with the nonlinear element. The angle between the wires is 49 deg.}
%%%%%%%%%%%%%%%%%

Figures~\rpict{properties} (a-c) show circular birefringence and polarization rotation for different incident intensities. The rotatory power of our sample exceeds that obtained earlier in an artificial structure~\cite{Rogacheva:2006-177401:PRL}, and it is five orders of magnitude stronger than the rotatory power of natural gyrotropic quartz crystal for optical wavelengths. The resonant feature observed for the circular dichroism comes from the resonant excitation of currents in the metallic wires of the left-handed metamolecule by the left-handed circularly polarized wave. Our numerical simulations confirm that the excited resonance corresponds to the out-of-phase currents in the wires. At the same time, the right-handed circularly polarized wave does not noticeably excite any resonances in our structure.

Changing the power of the incident wave shifts the resonance of the gyrotropic response to a higher frequency, see Fig.~\rpict{properties} (a-c). Importantly, such a shift of the polarization rotation resonance leads to {\em giant nonlinear gyrotropy} - see Fig.~\rpict{properties} (d-f). The nonlinear gyrotropy coefficient is calculated as NG=$\Delta\Theta/\Delta I /h$ deg$\times$cm/W, where $\Delta\Theta$ is the change in polarization rotation caused by the intensity change $\Delta I$, and $h=3.2$ mm is the sample thickness. We see that close to the resonant frequencies, the nonlinear gyrotropy reaches the value of 15 deg$\cdot$cm/W. There is no previous data available for microwave frequencies that we can compare our results to, but in natural optical media, e.g., in LiIO$_3$, the electronic nonlinear optical activity is 12 orders of magnitude weaker~\cite{Akhmanov:1979-294:JETPL,Polarization_of_Light}.

%%%%%%%%%%%%%%%%%
\pict[0.5]{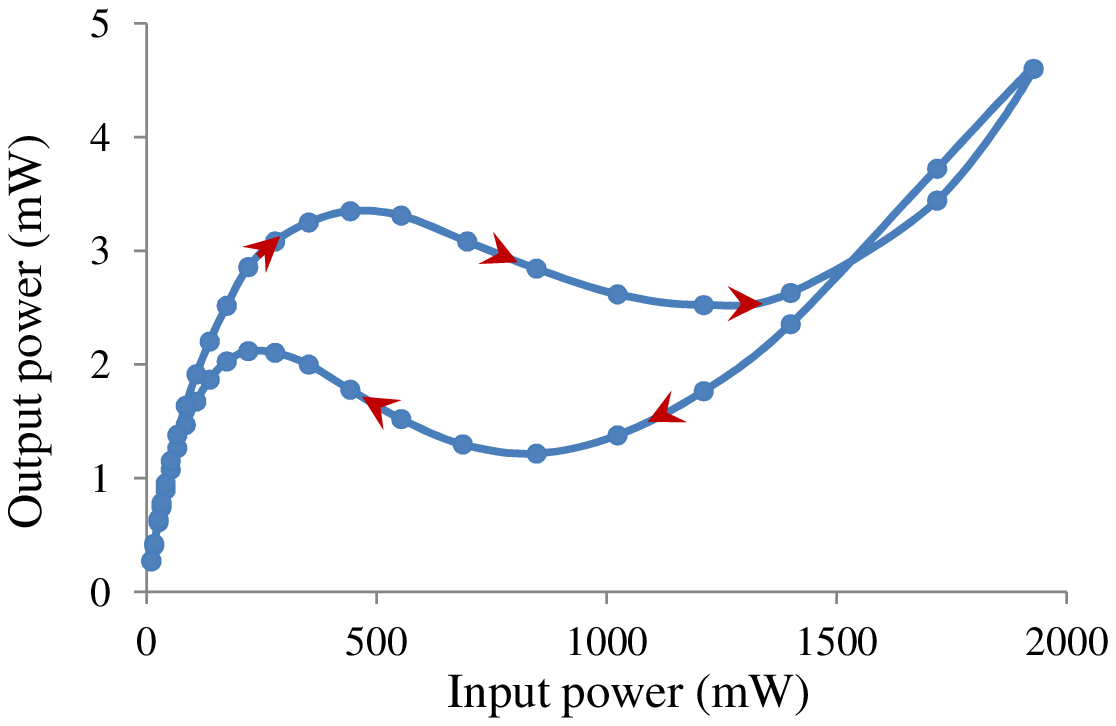}{bistability}{{\bf Bistable response of the nonlinear metamolecule.} Characteristic bistability and hysteresis response of the chiral metamolecule composed of two nonlinear wires illuminated by a left-handed circularly polarized wave of frequency 6.11 GHz. Each experimental point is obtained in the CW regime after the system reached the steady state.}
%%%%%%%%%%%%%%%%%

One of the important features of a nonlinear resonant system is the possibility of a bistable response. In order to study this effect, we use the continuous wave regime of the vector network analyzer. We selected the frequency of excitation to be 6.11 GHz, near the dip of the transmission curve obtained for the large intensity regime. We performed the measurements by first increasing the input power and measuring the steady state transmission amplitude for several values of input intensity (Fig.~\rpict{bistability}, top branch) and then similarly decreasing the power (Fig.~\rpict{bistability}, lower branch). We observe that the transmitted power may take different values which vary by a factor of three, depending on the history of the excitation.

%In conclusion, we report an optical diode which is a true analogue of an electric diode. This is a device transmitting electromagnetic waves in one direction only. Its functionality is underpinned by a truly giant nonlinear wave propagation polarization phenomenon in an artificial meta-molecule.

\begin{acknowledgments}
This work was supported by the Australian Research Council, by the Australian Academy of Science Travel Grant, and by the Engineering and Physical Sciences Research Council UK and The Royal Society (London).
\end{acknowledgments}

\vspace{0.5 true cm}

\begin{center}
{\bf METHODS}
\end{center}

{\em Structure.} Copper strips forming the chiral metamolecules were engraved on one side of 1.6 mm thick standard PCB laminate boards, as schematically shown in the inset to Fig. 2(a). The boards were circular to fit conveniently into the circular waveguide of the polarimeter. The wire pairs were formed by stacking two boards together such that the overall separation between the wires was 3.2 mm.  The metal strips had the following dimensions: width = 1 mm, length = 15 mm and thickness = 30 $\mu m$. We denote the metamolecule shown in the inset of Fig.~\rpict{diode} as a left-handed metamolecule, whereas its mirror-symmetric metamolecule is denoted as right-handed. The nonlinearity was added to the metamolecules by inserting a semiconductor varactor diode (model Skyworks SMV-1405), see Fig.~\rpict{diode}, in the middle of either one (asymmetric metamolecule) or both (symmetric metamolecule) wires. To match the resonance frequency of the varactor-loaded wires with that of the unloaded ones (which changes due to the internal capacitance of the varactor) the former were shortened to 14.3 mm. The introduction of the varactor also decreased the quality factor of the resonances due to its internal resistivity. A similar approach based on the use of a varactor diode was previously employed for creating nonlinear metamaterials~\cite{Shadrivov:2008-161903:APL,Powell:2009-084102:APL} as well as for tunability of epsilon-near-zero structures~\cite{Powell:2009-245135:PRB}.

{\em Measurements.} Experiments were performed in the 5.5--7.0 GHz frequency range using a Vector Network Analyzer (VNA, Agilent model E8364B) and a microwave waveguide polarimeter hosting the sample. The polarimeter was built around circular waveguide sections producing and guiding left-handed circularly polarized waves  (series 64 by Flann Microwave). In order to provide sufficient microwave power to observe nonlinear effects, we used an external  power amplifier (HP 83020A). To eliminate the contribution of the microwave cables and the waveguides, the measurements were normalized to the transmission coefficient of the empty waveguide.  To suppress the effect of multiple reflections from the waveguide components for the measurements of spectra of nonlinear structures, we use time-gating with 15 nsec gating window; this allows obtaining much smoother experimental curves. To measure the wave intensity fed into the waveguide, we use a 10 dB directional coupler connected to the input port of the waveguide. The power is measured with a Rohde and Schwartz power sensor, model Z23.

Similarly to Ref.~\cite{Rogacheva:2006-177401:PRL}, all our measurements were performed using left-handed circularly polarized waves only. The excitation of the samples with a right-handed circularly polarized wave is equivalent to the excitation of the corresponding mirror-reflected structures with a left-handed circularly polarized wave.

\end{sloppy}
\end{document}